\documentclass[aps,showpacs,nofootinbib,superscriptaddress,twocolumn,floatfix,amsmath,amssymb,amsfonts]{revtex4-1}

\usepackage{graphicx}
\usepackage{amssymb}
\usepackage{amsmath}
\usepackage{bm}
\usepackage{slashed}
\usepackage{datetime}
\usepackage{mciteplus}
\usepackage[utf8]{inputenc}
\usepackage{amsmath,amssymb,amsfonts}
\usepackage[svgnames]{xcolor}
\usepackage{natbib,hyperref}  

\begin{document}
\allowdisplaybreaks[2]

\title{First Extraction of Transversity from a Global Analysis of Electron-Proton \\ 
and Proton-Proton Data
}

\author{Marco Radici}
\email{marco.radici@pv.infn.it}
\affiliation{INFN Sezione di Pavia, via Bassi 6, I-27100 Pavia, Italy}

\author{Alessandro Bacchetta}
\email{alessandro.bacchetta@unipv.it}
\affiliation{Dipartimento di Fisica, Universit\`a di Pavia, via Bassi 6, I-27100 Pavia, Italy}
\affiliation{INFN Sezione di Pavia, via Bassi 6, I-27100 Pavia, Italy}

\begin{abstract}
We present the first extraction of the transversity distribution in the framework of collinear factorization based on the global analysis of pion-pair production in deep-inelastic scattering and in proton-proton collisions with a transversely polarized proton. The extraction relies on the knowledge of di-hadron fragmentation functions, which are taken from the analysis of electron-positron annihilation data. For the first time, the transversity is extracted from a global analysis similar to what is usually done for the spin-averaged and helicity distributions. The knowledge of transversity is important for, among other things, detecting possible signals of new physics in high-precision low-energy experiments. 
\end{abstract}

\date{\today, \currenttime}


\maketitle


Parton Distribution Functions (PDFs) describe the internal structure of hadrons in terms of number densities of confined quarks and gluons. If parton transverse momenta are integrated over, the complete spin structure of the nucleon at leading twist is described in terms of three PDFs: the spin-averaged PDF $f_1$, the helicity PDF $g_1$, and the transversity PDF $h_1$. The knowledge of PDFs is crucial for the interpretation of high-energy experiments involving hadrons and for detecting signals of new physics beyond the Standard Model (BSM). Transversity has recently received increasing attention because of the importance of a precise  determination of its integral, the so-called tensor charge $\delta q$ (for a flavor $q$)~\cite{Courtoy:2015haa}. In neutron $\beta$-decays, BSM effects can arise from the interference between SM operators and a new possible tensor operator whose coupling involves $\delta q$~\cite{Bhattacharya:2011qm}. Similarly, $\delta q$ enters the expression of the fermionic electric dipole moment that could constrain new possible CP-violating couplings in some BSM theories~\cite{Dubbers:2011ns,Yamanaka:2017mef}. In this Letter, we present an extraction of transversity at leading order (LO) in the strong coupling constant $\alpha_s$. For the first time we perform a global analysis of data on deep-inelastic scattering (DIS) and proton-proton collisions, {\it i.e.} for the first time we make a global analysis similar to what is usually done for the spin-averaged and the helicity distributions. 

The PDFs $f_1$ and $g_1$ are nowadays fairly well determined (see, e.g., Refs.~\cite{Butterworth:2015oua,Accardi:2016ndt} and references therein). On the contrary, we have a limited knowledge on $h_1$. Transversity describes the correlation between the transverse polarization of the nucleon and the transverse polarization of its constituent partons. The chiral-odd nature of $h_1$ makes it hard to extract it from experimental data because in the cross section transversity must always be paired to another chiral-odd object~\cite{Jaffe:1991kp}. Transversity was extracted for the first time from data on single-particle semi-inclusive DIS (SIDIS)~\cite{Anselmino:2007fs} where the cross section contains an azimuthal modulation proportional to the convolution $h_1 \otimes H_1^\perp$ involving the chiral-odd Collins fragmentation function $H_1^\perp$~\cite{Collins:1993kk}. However, the convolution $h_1 \otimes H_1^\perp$ involves the  transverse momenta of quarks. Therefore, the evolution of the various partonic functions among different experimental energies must be described in the framework of transverse-momentum dependent factorization, which depends also on non-perturbative parameters~\cite{Kang:2015msa}. More importantly, this analysis of the so-called Collins effect cannot be extended to hadronic collisions because of factorization breaking contributions~\cite{Rogers:2010dm}. 

An alternative method to access the transversity PDF requires only standard collinear factorization, thus avoiding the above complications and limitations. It considers the semi-inclusive production of two hadrons with small invariant mass~\cite{Jaffe:1998hf}, and is based on the correlation between the transverse polarization of the quark fragmenting into the two hadrons and their transverse relative momentum~\cite{Collins:1994ax}. In this case, the di-hadron SIDIS cross section (once integrated over partonic transverse momenta) contains a specific modulation in the azimuthal orientation of the plane containing the momenta of the two hadrons. The coefficient of this modulation is the simple product $h_1 H_1^\sphericalangle$ where $H_1^\sphericalangle$ is a chiral-odd di-hadron fragmentation function (DiFF) quantifying the above correlation~\cite{Bianconi:1999cd,Radici:2001na,Bacchetta:2002ux}. The function $H_1^\sphericalangle$ can be independently determined by looking at correlations between the azimuthal orientations of two hadron pairs in back-to-back jets in $e^+ e^-$ annihilation~\cite{Boer:2003ya,Bacchetta:2008wb,Courtoy:2012ry}. The advantage of this method with respect to the Collins effect is that collinear factorization makes it possible to isolate the same combination $h_1 H_1^\sphericalangle$ also in proton-proton collisions~\cite{Bacchetta:2004it}, giving rise to an azimuthally asymmetric distribution of the final hadron pair when one of the two initial protons is transversely polarized~\cite{Radici:2016lam}. 

Experimental data for the SIDIS asymmetry in the azimuthal distribution of final $(\pi^+ \pi^-)$ pairs were first collected by the {\tt HERMES} collaboration for a transversely polarized proton target~\cite{Airapetian:2008sk}, and by the {\tt COMPASS} collaboration for polarized protons and deuterons~\cite{Adolph:2012nw,Adolph:2014fjw,Braun:2015baa}. The azimuthal asymmetry in the distribution of back-to-back $(\pi^+ \pi^-)$ pairs in $e^+ e^-$ annihilation was measured by the {\tt BELLE} collaboration~\cite{Vossen:2011fk}, opening the way to the first parametrization of $H_1^\sphericalangle$ for the up and down quarks~\cite{Courtoy:2012ry}. This result was used in combination with the SIDIS data to extract the valence components of $h_1$~\cite{Bacchetta:2011ip,Bacchetta:2012ty,Radici:2015mwa}. Recently, the {\tt STAR} collaboration released the first results for the relevant asymmetry in the azimuthal distribution of $(\pi^+ \pi^-)$ pairs produced in proton-proton collisions with a transversely polarized proton~\cite{Adamczyk:2015hri}. Here, we present for the first time the extraction of the transversity PDF $h_1$ from a global fit of all these data.


For the SIDIS process $e[k] + A^\uparrow [P] \rightarrow e' [k'] + (\pi^+ \pi^-)[P_h] + X$, where an electron with 4-momentum $k$ scatters off a transversely polarized proton $(A=p)$ or deuteron $(A=d)$ with 4-momentum $P$ at the hard scale $Q^2 = -q^2 = -(k - k')^2 \geq 0$ producing a $(\pi^+ \pi^-)$ pair with total 4-momentum $P_h$ and relative 4-momentum $R$ plus any number of undetected hadrons $(X)$, the relevant asymmetry at leading twist (usually denoted as $A_{\mathrm{UT}}^{\sin (\phi_R+\phi_S) \sin\theta}$~\cite{Bacchetta:2011ip,Bacchetta:2012ty,Radici:2015mwa}) can be written as
\begin{equation}
A_{\mathrm{DIS}} (x, Q^2) = - C_y \, \frac{\sum_q \, e_q^2 \, h_1^q (x, Q^2) \, n_q^\uparrow (Q^2)}{\sum_q \, e_q^2 \, f_1^q (x, Q^2) \, n_q (Q^2)} \, ,
\label{e:Adis}
\end{equation}
where $x = Q^2 / (2 P \cdot q)$ is the fractional momentum of the initial quark, $C_y$ is a coefficient depending  on the average value of the fractional beam energy loss $y = P \cdot q / (P \cdot k)$, $e_q$ is the fractional electric charge of a quark $q$, and 
\begin{align}
n_q (Q^2) &= \int dz \int dM_h \, D_1^q (z, M_h, Q^2) \nonumber \\
n_q^\uparrow (Q^2) &= \int dz \int dM_h \, \frac{|\bm{R} |}{M_h} \, H_1^{\sphericalangle\, q} (z, M_h, Q^2) \, , 
\label{e:intDiFF}
\end{align}
where $z = P\cdot P_h / (P \cdot q) = z_{\pi^+} + z_{\pi^-}$ is the fractional energy carried by the $(\pi^+ \pi^-)$ pair, $M_h$ is its invariant mass $(M_h^2 = P_h^2 \ll Q^2)$, and the DiFFs $D_1^q$ and $H_1^{\sphericalangle \, q}$ describe the fragmentation into the $(\pi^+ \pi^-)$ pair of an unpolarized or transversely polarized quark $q$, respectively. Data was collected in bins of $x, \, z,$ and $M_h,$ for $(\pi^+ \pi^-)$ pairs and for final unidentified $(h^+ h^-)$ pairs~\cite{Adolph:2012nw,Adolph:2014fjw}. Since our goal is to extract transversity from data for the inclusive $(\pi^+ \pi^-)$ production, here we consider the data set for identified $(\pi^+ \pi^-)$ pairs with only the bins in $x$~\cite{Airapetian:2008sk,Braun:2015baa}, whose average value spans the range $0.0065 \leq \langle x \rangle \leq 0.133$ corresponding to the average scale range $1.232 \leq \langle Q^2 \rangle \leq 31.5$ GeV$^2$. 

The unknown factors $n_q, \, n_q^\uparrow$ of Eq.~\eqref{e:intDiFF} can be inferred from the process $e^+[\bar{k}] + e^-[k] \rightarrow (\pi^+ \pi^-)[P_h] + (\pi^+ \pi^-)[\bar{P}_h] + X$ at the hard scale $Q^2 = (k + \bar{k})^2 \geq 0$, where the two $(\pi^+ \pi^-)$ pairs are emitted in opposite hemispheres (ensured by the condition $P_h \cdot \bar{P}_h \approx Q^2$). By summing the $(\pi^+ \pi^-)$ pairs of one hemisphere, the so-called Artru-Collins asymmetry (usually denoted as $A^{\cos (\phi_R+\phi_{\bar{R}})}$~\cite{Boer:2003ya,Bacchetta:2008wb,Courtoy:2012ry,Matevosyan:2018icf,Artru:1995zu} or $a_{12R}$ in the {\tt BELLE} publication~\cite{Vossen:2011fk}) is given by 
\begin{align}
A_{e^+ e^-} (z, M_h, Q^2) &= - C_{\theta_2} \, \frac{|\bm{R}|}{M_h} \nonumber \\
&\hspace{-0.5cm} \times \frac{\sum_q \, e_q^2 \, H_1^{\sphericalangle \, q} (z, M_h, Q^2) \, n_q^\uparrow (Q^2)}{\sum_q \, e_q^2 \, D_1^q (z, M_h, Q^2) \, n_q (Q^2)} \, ,
\label{e:Ae+e-}
\end{align}
where $C_{\theta_2}$ is a coefficient depending on the average value of the angle $\theta_2$ between the $\bar{\bm{k}}$ and $\bm{P}_h$ directions. The $D_1^q$ is parametrized to reproduce the two-pion yield of the {\tt PYTHIA} event generator tuned to the {\tt BELLE} kinematics~\cite{Courtoy:2012ry}. Data span the range $0.2\leq z \leq 1$ and $0.3 \leq M_h \leq 1.2$ GeV at $Q=10$ GeV. (When this Letter was being finalized, the {\tt BELLE} collaboration has officially published the first data for the differential di-hadron multiplicities~\cite{Seidl:2017qhp}. We will use these data in a future work to parametrize the unpolarized DiFF  $D_1$ directly from experiment.)  

The same elementary mechanism $h_1 H_1^\sphericalangle$ active in SIDIS generates an azimuthal  asymmetry also in the $p[P_A] + p^\uparrow[P_B] \rightarrow (\pi^+ \pi^-)[P_h] + X$ process~\cite{Bacchetta:2004it}, where a proton with 4-momentum $P_A$ collides with a transversely polarized proton with 4-momentum $P_B$. After integrating over the partonic transverse momenta, the total 4-momentum $P_h$ of the $(\pi^+ \pi^-)$ pair has no transverse component with respect to the fragmenting quark momentum but it can have the transverse component $\bm{P}_{hT}$ with respect to $P_A$. The $\bm{P}_{hT}^2$ represents the hard scale of the process $(\bm{P}_{hT}^2 \gg M_h^2 = P_h^2)$. If we identify the reaction plane by $(\bm{P}_A, \bm{P}_h)$, the relevant asymmetry in the azimuthal distribution of $(\pi^+ \pi^-)$ pairs with respect to the reaction plane (usually denoted as $A_{UT}$~\cite{Adamczyk:2015hri,Radici:2016lam}) is given at leading twist by
\begin{equation}
A_{pp} (\eta, |\bm{P}_{hT}|, M_h) = \frac{\pi}{4}\, \frac{|\bm{R}|}{M_h} \, \frac{H (\eta, |\bm{P}_{hT}|, M_h)}{D (\eta, |\bm{P}_{hT}|, M_h)} \, , 
\label{e:App}
\end{equation}
where $\eta$ is the pseudorapidity of the hadron pair with respect to the beam $\bm{P}_A$, and
\begin{align}
H (\eta, |\bm{P}_{hT}|, M_h) &= \sum_{a,b,c,d}  \int \frac{dx_a \, dx_b}{z_h} \, f_1^a (x_a) \, h_1^b (x_b) \nonumber \\
&\times\, \frac{d\Delta \sigma_{a b^\uparrow \rightarrow c^\uparrow d}}{d \hat{t}} \, H_1^{\sphericalangle\, c} (z_h, M_h) \, , \label{e:numApp} \\
D (\eta, |\bm{P}_{hT}|, M_h) &=  \sum_{a,b,c,d} \int \frac{dx_a \, dx_b}{z_h} \, f_1^a (x_a) \, f_1^b (x_b) \nonumber \\
& \times   \, \frac{d\sigma_{a b \rightarrow c d}}{d \hat{t}} \, D_1^c (z_h, M_h)  \, . \label{e:denApp}
\end{align}
The dependence upon the hard scale $\bm{P}_{hT}^2$ is understood in all PDFs and DiFFs. The elementary annihilation of partons $a$ and $b$ (carrying fractional momenta $x_a$ and $x_b$, respectively) into the partons $c$ and $d$ is described by the cross section $d\sigma$, while $d\Delta\sigma$ refers to the transfer of transverse polarization in the same mechanism~\cite{Bacchetta:2004it}. Both cross sections are differential in $\hat{t} = t \, x_a / z_h$, where $t = (P_A - P_B)^2$ and $z_h$ is the fractional energy carried by the pion pair, which is related by momentum conservation to $\bm{P}_{hT}^2, \, \eta, \, x_a, \, x_b,$ and $s = (P_A + P_B)^2$ (the squared center-of-mass energy in the collision)~\cite{Bacchetta:2004it}. Data for $A_{pp}$ were collected by the {\tt STAR} collaboration at $\sqrt{s} = 200$ GeV~\cite{Adamczyk:2015hri} in bins of $\eta, \, |\bm{P}_{hT}|$, and $M_h$, after integrating on the complementary variables. The average values are limited to the ranges $-0.84 \leq \eta \leq 0.84$, $3 \leq |\bm{P}_{hT}| \leq 13$ GeV, and $0.3 \leq M_h \leq 1.2$ GeV, which correspond to $0.15 \lesssim \langle x \rangle \lesssim 0.3$ in SIDIS but at a larger hard scale.


In order to reduce the computational time, following Ref.~\cite{Stratmann:2001pb} we rewrite the parameter-dependent part of Eq.~\eqref{e:numApp} in Mellin space (see~\cite{suppl}). In order to exploit this workaround, it is crucial that the Mellin transform of $h_1$ can be analytically calculated at any scale. The functional form adopted in previous fits of di-hadron SIDIS data does not match this criterion~\cite{Bacchetta:2012ty,Radici:2015mwa}. Here, we have modified it, but kept its main features: a) satisfying the Soffer inequality at any scale $Q^2$; b) having a high degree of flexibility with up to three nodes in $x$. Since the Soffer bound is valid for each quark and antiquark and we need to parametrize their valence combination $q_v = q-\bar{q}$, we constrain the transversity by taking the sum of Soffer bounds for both quarks and antiquarks~\cite{Bacchetta:2012ty}. At $Q_0^2 = 1$ GeV$^2$, the general structure of the functional form is given by
\begin{equation}
x\, h_1^{q_v} (x, Q_0^2) = F^q (x) \, F_{\mathrm{SB}}^q (x) \, , 
\label{e:h1xQ0}
\end{equation}
where $F_{\mathrm{SB}}^q (x)$ is a fit to the sum of the Soffer bounds for $q$ and $\bar{q}$ at $Q_0^2$~\cite{suppl} and 
\begin{align}
F^q (x) &= N_F^q \, \frac{{\cal F}^q (x)}{\mathrm{max}_x [|{\cal F}^q (x)|]} \, , \label{e:h1xQ0_F} \\
{\cal F}^q (x) &= x^{A_q} \, \left[ 1 + B_q \, T_1(x) + C_q \, T_2 (x) + D_q \, T_3 (x) \right]  \, . \nonumber
\end{align}
The $T_n (x)$ are the Cebyshev polynomials of order $n$. The $N_F^q, \, A_q, \, B_q, \, C_q, \, D_q, $ are fitting parameters. Simplifying assumptions on the isospin symmetry of DiFFs allow us to access only the valence components of transversity~\cite{Bacchetta:2006un,Bacchetta:2011ip}; hence, we have in total 10 free parameters. If we impose the constraint $|N_F^q | \leq 1$, then $|F^q (x)| \leq 1$ for all $x$, and the transversity of Eq.~\eqref{e:h1xQ0} automatically satisfies the Soffer inequality at any scale~\cite{Vogelsang:1997ak}. In this Letter, we keep for $f_1$ and $g_1$ the same parametrizations as in our previous fits (MSTW08 at LO for $f_1$~\cite{Martin:2009iq} and DSSV for $g_1$~\cite{deFlorian:2009vb}). [We checked that replacing DSSV with more recent parametrizations of $g_1$ (like JAM15~\cite{Sato:2016tuz} or JAM17~\cite{Ethier:2017zbq}) does not make any relevant change to our results within the current experimental and theoretical uncertainties.] With this choice, the Soffer bound at $Q_0^2$ can be reproduced by the $F_{\mathrm{SB}}$ at 1\% accuracy in the range $0.001\leq x \leq 1$~\cite{suppl}. From Eqs.~\eqref{e:h1xQ0} and~\eqref{e:h1xQ0_F}, we deduce that $x h_1^{q_v}(x) \approx x^{A_q + a_q}$ at very small $x$. This asymptotic behavior is strongly constrained by requiring that the tensor charge 
\begin{equation}
\delta q (Q^2) = \int_0^1 dx \, h_1^{q_v}(x,Q^2) 
\label{e:gT}
\end{equation} 
is finite. We numerically evaluate the integral in the range $[x_{\mathrm{min}}, 1]$ where for MSTW08 $x_{\mathrm{min}} = 10^{-6}$~\cite{Martin:2009iq}. In order to avoid uncontrolled extrapolation errors below  $x_{\mathrm{min}}$, we impose the condition $A_q + a_q > 1/3$, which also grants that $\delta q$ is evaluated at 1\% accuracy. (According to Ref.~\cite{Accardi:2017pmi}, the more stringent condition $A_q + a_q > 1$ is required to avoid a violation of the Burkardt-Cottingham sum rule by an infinite amount.) 

By inserting Eq.~\eqref{e:h1xQ0_F} in Eq.~\eqref{e:h1xQ0} and using the function $F_{\mathrm{SB}}$ listed in~\cite{suppl}, the resulting expression can be easily transformed in Mellin space and evolved at LO~\cite{suppl}. When dealing with the $A_{\mathrm{DIS}}$ of Eq.~\eqref{e:Adis}, the $h_1(x,Q_0^2)$ of Eq.~\eqref{e:h1xQ0} is evolved using the {\tt HOPPET} code~\cite{Salam:2008qg} suitably extended to include LO chiral-odd splitting functions (and, similarly, for $H_1^\sphericalangle$~\cite{Ceccopieri:2007ip}).  

The statistical uncertainty of the global fit is studied using the same bootstrap method as in our previous fits~\cite{Bacchetta:2012ty,Radici:2015mwa}. In the following, for a set of $M$ replicas of the data points the statistical error is constructed by taking the central 90\% of them, namely by rejecting the largest and smallest 5\% of the $M$ replicas for each experimental bin. The theoretical result is obtained by integrating the asymmetry over the bin width of the displayed variable, after integrating over the full range of the other ones. In the analysis of di-hadron $e^+ e^-$ data, the $D_1^q$ with $q=u, d, s, c,$ is parametrized from the {\tt PYTHIA} yield at the {\tt BELLE} kinematics~\cite{Courtoy:2012ry} assuming that $D_1^g (z, M_h; Q_0^2) = 0$. We parametrize the error on the unconstrained $D_1^g$ by computing the denominator $D$ in Eq.~\eqref{e:denApp} alternatively with $D_1^g (z, M_h; Q_0^2) = D_1^u (z, M_h; Q_0^2) / 4$, or $D_1^u (z, M_h; Q_0^2)$. We have verified that these choices alter the $\chi^2$ of the $e^+ e^-$ fit in Ref.~\cite{Courtoy:2012ry} by 10-50\%, keeping always $\chi^2$/d.o.f. $\lesssim 2$ . The number $M$ of replicas is fixed by reproducing the mean and standard deviation of the original data points. For each option, it turns out that 200 replicas are sufficient. Hence, we have in total $M=600$ replicas.


For a total of 46 bins and 10 free parameters, we reach a global $\chi^2$/d.o.f. $= 2.08 \pm 0.09$. The SIDIS data contribute to the global $\chi^2$ by $\approx 38$\%, most of which ($\approx 76$\%) coming from {\tt COMPASS} data points, due to their smaller errors. A significant amount ($\approx 40$\%) is contributed to the {\tt COMPASS} $\chi^2$ budget by specific bins in the deuteron kinematics. The remaining 62\% of the global $\chi^2$ comes from the {\tt STAR} data and is dominated by the $|\bm{P}_{hT}|$ bins ($\approx 70$\%),  while the $M_h$ bins contribute by $\approx 28$\% and the $\eta$ bins by a negligible $\approx 2$\%.

\begin{figure}
\begin{center}
\includegraphics[width=0.35\textwidth]{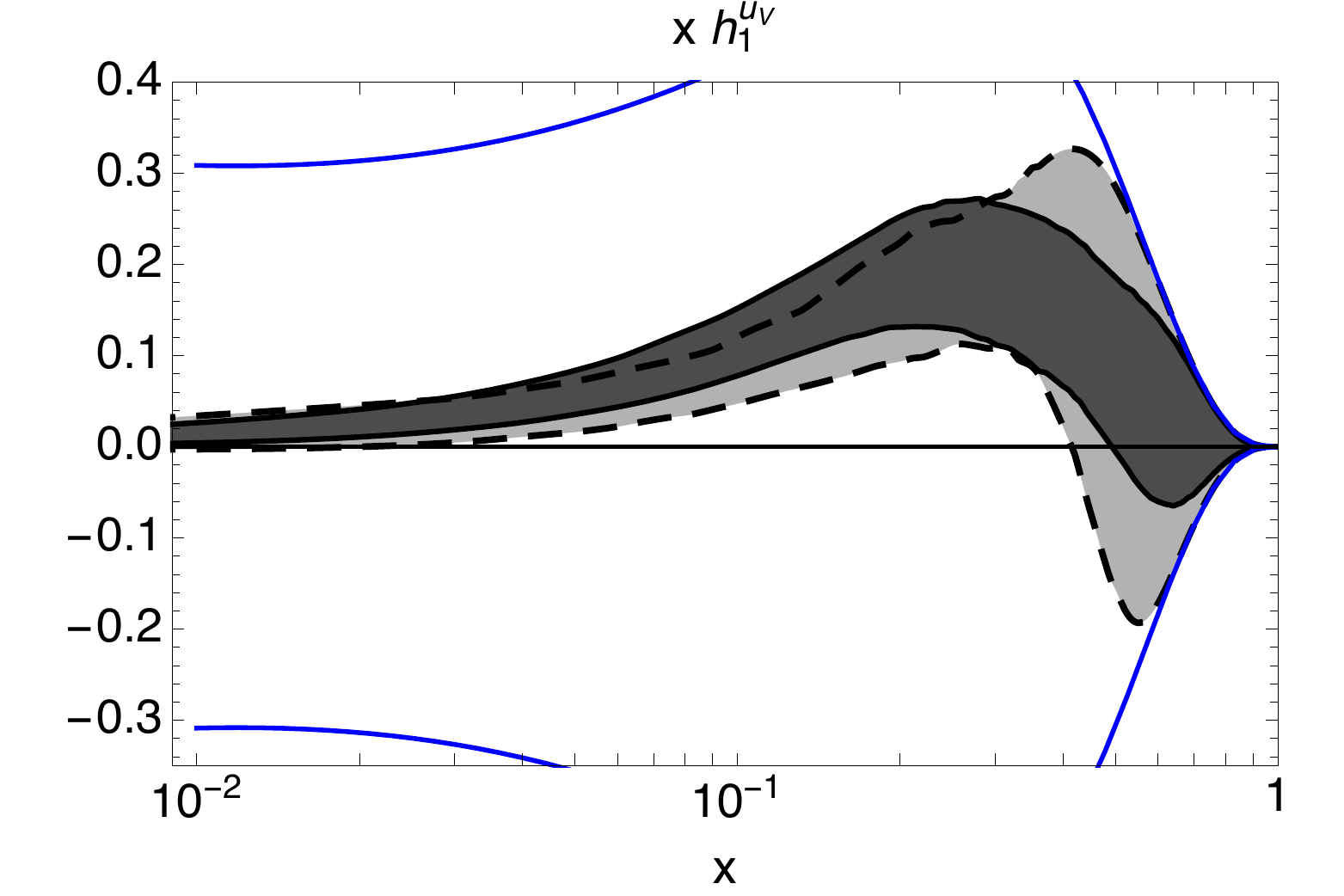} \\
\includegraphics[width=0.35\textwidth]{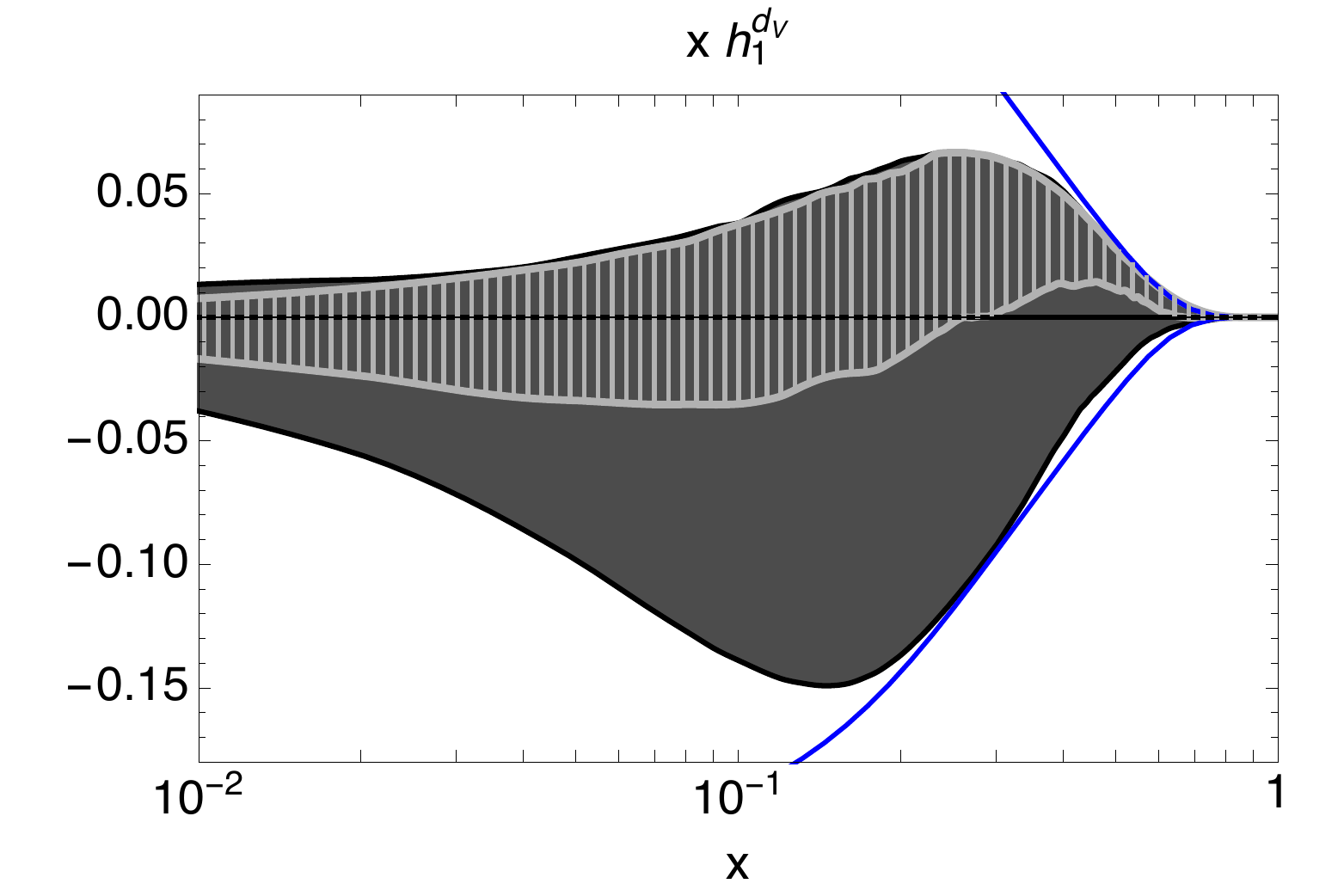} 
\end{center}
\caption{The transversity $x\, h_1$ as a function of $x$ at $Q^2 = 2.4$ GeV$^2$. Dark (blue) lines represent the Soffer bounds. Dark bands with solid borders for the global fit of this work including all options $D_1^g (Q_0^2) = 0$, $D_1^u (Q_0^2) / 4$ and $D_1^u (Q_0^2)$. (Top) For valence up quark: comparison with our previous fit in Ref.~\cite{Radici:2015mwa} (lighter band with dashed borders). (Bottom) For valence down quark: comparison with this global fit with only $D_1^g (Q_0^2) = 0$ (hatched area with lighter borders).}
\label{f:xh1}
\end{figure}

In Fig.~\ref{f:xh1}, the transversity $x\, h_1$ is displayed as a function of $x$ at $Q^2 = 2.4$ GeV$^2$. The dark (blue) lines represent the Soffer bounds. The upper panel refers to the valence up component. Here, the lighter band with dashed borders corresponds to the 90\% uncertainty band from our previous fit with only SIDIS and $e^+ e^-$ data~\cite{Radici:2015mwa}. The darker band with solid borders is the 90\% uncertainty band from the new global fit discussed here, including all options $D_1^g (Q_0^2) = 0$, $D_1^g (Q_0^2) = D_1^u (Q_0^2) / 4$ and $D_1^g (Q_0^2) = D_1^u (Q_0^2)$. However, the latter result is insensitive to the various choices for $D_1^g (Q_0^2)$. There is an evident gain in precision by including also the {\tt STAR} data. The uncertainty of our previous fit in Ref.~\cite{Radici:2015mwa} (the lighter band) is comparable to the one obtained from the analysis of the Collins effect~\cite{Kang:2015msa,Anselmino:2013vqa}. Hence, we deduce that the outcome of our global fit provides a substantial increase in the precision on $h_1^{u_v}$ and on the related tensor charge $\delta u$ with respect to all the other phenomenological extractions.


\begin{table}[ht]
\begin{tabular}{c c c c c}
\hline\hline\\[-4.5mm]
                                 & $\delta q$ & $\delta q$ & $\delta \tilde{q}$ & $g_T$  \\ \hline \\[-4.5mm]
$(Q^2[$ GeV$^2])$ & $Q_0^2=1$ & $Q^2=4$ & $Q^2=10$ &$Q^2=4$   \\
\hline
\begin{tabular}{c}
up \\  down
\end{tabular} 
& 
\begin{tabular}{c}
$0.43(11)$ \\
$- 0.12(28)$
\end{tabular}
&
\begin{tabular}{c}
$0.39(10)$ \\
$- 0.11(26)$
\end{tabular}
&
\begin{tabular}{c}
$0.32(8)$   \\
$- 0.10(22)$  
\end{tabular}
&
$0.53(25)$ \\
\hline\hline
\end{tabular}
\caption{The tensor charge $\delta q$, truncated tensor charge $\delta \tilde{q}$, and isovector tensor charge $g_T$ at 90\% confidence level (see text).}
\label{t:gT}
\end{table}


The lower panel of Fig.~\ref{f:xh1} refers to the valence down component. Again, the darker band with solid borders corresponds to the 90\% uncertainty band from the new global fit including all options for $D_1^g (Q_0^2)$. The hatched area with lighter borders shows how the result is modified by including only the option $D_1^g (Q_0^2) = 0$. At variance with the up quark, the valence down component $h_1^{d_v}$ is sensitive to the $D_1^g$ contribution to the cross section for $p$-$p$ collisions. Data on $(\pi^+ \pi^-)$ multiplicities in $p$-$p$ collisions would be very useful in constraining $D_1^g$. Finally, we notice that the unnatural behavior of $h_1^{d_v}$ at $x \gtrsim 0.1$ obtained in Ref.~\cite{Radici:2015mwa} has disappeared. The few {\tt COMPASS} data points responsible for this anomalous trend~\cite{Radici:2016opu} become statistically less relevant when including the {\tt STAR} data, which in turn demonstrate their large impact on our knowledge of transversity.

In order to compare with other results, we have calculated at various scales the tensor charge $\delta q$ and the truncated tensor charge $\delta \tilde{q}$, which is obtained by restricting the integral in Eq.~\eqref{e:gT} to the experimental $x$ range [0.0065, 0.133]. The values at 90\% confidence level are listed in Tab.~\ref{t:gT}. They are in very good agreement with other phenomenological extractions~\cite{Kang:2015msa}, and seem compatible with lattice simulations for $\delta d$ but not for $\delta u$ (see, e.g., Table IX in Ref.~\cite{Bhattacharya:2015wna}). We have computed also the isovector tensor charge $g_T \equiv \delta u - \delta d$, whose systematic errors of lattice calculations are under better control. At $Q^2 = 4$ GeV$^2$, our result is again in very good agreement with phenomenology (displaying a significantly smaller error with respect to our previous fit with only di-hadron SIDIS and $e^+ e^-$ data~\cite{Radici:2015mwa}) but seems incompatible with most recent lattice simulations~\cite{Kang:2015msa,Bhattacharya:2015wna}. Lately, the authors of Ref.~\cite{Lin:2017stx} have published a Monte Carlo re--analysis of the Collins effect supplemented by lattice input for $g_T$, showing that their result for $g_T$ is compatible with some lattice calculations although their numerical values for $\delta u$ and $\delta d$ turn out to be largely incompatible. More work is needed along both lines of improving the precision of phenomenological extractions and of benchmarking lattice simulations, because a careful determination of $g_T$ is of paramount importance in detecting BSM effects~\cite{Courtoy:2015haa}, e.g., in neutron $\beta$-decay, where the experimental accuracy has now reached the 0.1\% level~\cite{Cirigliano:2013xha}. 

In summary, we have presented an extraction of transversity at leading order in the strong coupling constant where, for the first time, we performed a global fit of all data for azimuthal asymmetries in the semi-inclusive production of charged pion pairs in deep-inelastic scattering, electron-positron annihilation, and proton-proton collisions. 
The uncertainty on the result for the valence up quark is smaller than any previous extraction. The large sensitivity of the valence down quark to the unconstrained gluon channel in di-hadron fragmentation calls for data on pion pair multiplicities in proton-proton collisions, which are still missing. The calculated isovector tensor charge seems incompatible with most lattice calculations. 

M.R. thanks U. D'Alesio and G. Bozzi for useful discussions. This work is supported by the European Research Council (ERC) under the European Union's Horizon 2020 research and innovation program (grant agreement No. 647981, 3DSPIN). 


\bibliographystyle{apsrevM}
\bibliography{mybiblio}

\ifx\mcitethebibliography\mciteundefinedmacro
\PackageError{apsrevM.bst}{mciteplus.sty has not been loaded}
{This bibstyle requires the use of the mciteplus package.}\fi
\begin{mcitethebibliography}{49}
\expandafter\ifx\csname natexlab\endcsname\relax\def\natexlab#1{#1}\fi
\expandafter\ifx\csname bibnamefont\endcsname\relax
  \def\bibnamefont#1{#1}\fi
\expandafter\ifx\csname bibfnamefont\endcsname\relax
  \def\bibfnamefont#1{#1}\fi
\expandafter\ifx\csname citenamefont\endcsname\relax
  \def\citenamefont#1{#1}\fi
\expandafter\ifx\csname url\endcsname\relax
  \def\url#1{\texttt{#1}}\fi
\expandafter\ifx\csname urlprefix\endcsname\relax\def\urlprefix{URL }\fi
\providecommand{\bibinfo}[2]{#2}
\providecommand{\eprint}[2][]{\url{#2}}

\bibitem[{\citenamefont{Courtoy et~al.}(2015)\citenamefont{Courtoy, Baessler,
  Gonzalez-Alonso, and Liuti}}]{Courtoy:2015haa}
\bibinfo{author}{\bibfnamefont{A.}~\bibnamefont{Courtoy}},
  \bibinfo{author}{\bibfnamefont{S.}~\bibnamefont{Baessler}},
  \bibinfo{author}{\bibfnamefont{M.}~\bibnamefont{Gonzalez-Alonso}},
  \bibnamefont{and} \bibinfo{author}{\bibfnamefont{S.}~\bibnamefont{Liuti}},
  \bibinfo{journal}{Phys. Rev. Lett.} \textbf{\bibinfo{volume}{115}},
  \bibinfo{pages}{162001} (\bibinfo{year}{2015}), \eprint{1503.06814}\relax
\mciteBstWouldAddEndPuncttrue
\mciteSetBstMidEndSepPunct{\mcitedefaultmidpunct}
{\mcitedefaultendpunct}{\mcitedefaultseppunct}\relax
\EndOfBibitem
\bibitem[{\citenamefont{Bhattacharya et~al.}(2012)\citenamefont{Bhattacharya,
  Cirigliano, Cohen, Filipuzzi, Gonzalez-Alonso et~al.}}]{Bhattacharya:2011qm}
\bibinfo{author}{\bibfnamefont{T.}~\bibnamefont{Bhattacharya}},
  \bibinfo{author}{\bibfnamefont{V.}~\bibnamefont{Cirigliano}},
  \bibinfo{author}{\bibfnamefont{S.~D.} \bibnamefont{Cohen}},
  \bibinfo{author}{\bibfnamefont{A.}~\bibnamefont{Filipuzzi}},
  \bibinfo{author}{\bibfnamefont{M.}~\bibnamefont{Gonzalez-Alonso}},
  \bibnamefont{et~al.}, \bibinfo{journal}{Phys.Rev.}
  \textbf{\bibinfo{volume}{D85}}, \bibinfo{pages}{054512}
  (\bibinfo{year}{2012}), \eprint{1110.6448}\relax
\mciteBstWouldAddEndPuncttrue
\mciteSetBstMidEndSepPunct{\mcitedefaultmidpunct}
{\mcitedefaultendpunct}{\mcitedefaultseppunct}\relax
\EndOfBibitem
\bibitem[{\citenamefont{Dubbers and Schmidt}(2011)}]{Dubbers:2011ns}
\bibinfo{author}{\bibfnamefont{D.}~\bibnamefont{Dubbers}} \bibnamefont{and}
  \bibinfo{author}{\bibfnamefont{M.~G.} \bibnamefont{Schmidt}},
  \bibinfo{journal}{Rev. Mod. Phys.} \textbf{\bibinfo{volume}{83}},
  \bibinfo{pages}{1111} (\bibinfo{year}{2011}), \eprint{1105.3694}\relax
\mciteBstWouldAddEndPuncttrue
\mciteSetBstMidEndSepPunct{\mcitedefaultmidpunct}
{\mcitedefaultendpunct}{\mcitedefaultseppunct}\relax
\EndOfBibitem
\bibitem[{\citenamefont{Yamanaka et~al.}(2017)\citenamefont{Yamanaka, Sahoo,
  Yoshinaga, Sato, Asahi, and Das}}]{Yamanaka:2017mef}
\bibinfo{author}{\bibfnamefont{N.}~\bibnamefont{Yamanaka}},
  \bibinfo{author}{\bibfnamefont{B.~K.} \bibnamefont{Sahoo}},
  \bibinfo{author}{\bibfnamefont{N.}~\bibnamefont{Yoshinaga}},
  \bibinfo{author}{\bibfnamefont{T.}~\bibnamefont{Sato}},
  \bibinfo{author}{\bibfnamefont{K.}~\bibnamefont{Asahi}}, \bibnamefont{and}
  \bibinfo{author}{\bibfnamefont{B.~P.} \bibnamefont{Das}},
  \bibinfo{journal}{Eur. Phys. J.} \textbf{\bibinfo{volume}{A53}},
  \bibinfo{pages}{54} (\bibinfo{year}{2017}), \eprint{1703.01570}\relax
\mciteBstWouldAddEndPuncttrue
\mciteSetBstMidEndSepPunct{\mcitedefaultmidpunct}
{\mcitedefaultendpunct}{\mcitedefaultseppunct}\relax
\EndOfBibitem
\bibitem[{\citenamefont{Butterworth et~al.}(2016)}]{Butterworth:2015oua}
\bibinfo{author}{\bibfnamefont{J.}~\bibnamefont{Butterworth}}
  \bibnamefont{et~al.}, \bibinfo{journal}{J. Phys.}
  \textbf{\bibinfo{volume}{G43}}, \bibinfo{pages}{023001}
  (\bibinfo{year}{2016}), \eprint{1510.03865}\relax
\mciteBstWouldAddEndPuncttrue
\mciteSetBstMidEndSepPunct{\mcitedefaultmidpunct}
{\mcitedefaultendpunct}{\mcitedefaultseppunct}\relax
\EndOfBibitem
\bibitem[{\citenamefont{Accardi et~al.}(2016)}]{Accardi:2016ndt}
\bibinfo{author}{\bibfnamefont{A.}~\bibnamefont{Accardi}} \bibnamefont{et~al.},
  \bibinfo{journal}{Eur. Phys. J.} \textbf{\bibinfo{volume}{C76}},
  \bibinfo{pages}{471} (\bibinfo{year}{2016}), \eprint{1603.08906}\relax
\mciteBstWouldAddEndPuncttrue
\mciteSetBstMidEndSepPunct{\mcitedefaultmidpunct}
{\mcitedefaultendpunct}{\mcitedefaultseppunct}\relax
\EndOfBibitem
\bibitem[{\citenamefont{Jaffe and Ji}(1991)}]{Jaffe:1991kp}
\bibinfo{author}{\bibfnamefont{R.~L.} \bibnamefont{Jaffe}} \bibnamefont{and}
  \bibinfo{author}{\bibfnamefont{X.}~\bibnamefont{Ji}}, \bibinfo{journal}{Phys.
  Rev. Lett.} \textbf{\bibinfo{volume}{67}}, \bibinfo{pages}{552}
  (\bibinfo{year}{1991})\relax
\mciteBstWouldAddEndPuncttrue
\mciteSetBstMidEndSepPunct{\mcitedefaultmidpunct}
{\mcitedefaultendpunct}{\mcitedefaultseppunct}\relax
\EndOfBibitem
\bibitem[{\citenamefont{Anselmino et~al.}(2007)}]{Anselmino:2007fs}
\bibinfo{author}{\bibfnamefont{M.}~\bibnamefont{Anselmino}}
  \bibnamefont{et~al.}, \bibinfo{journal}{Phys. Rev.}
  \textbf{\bibinfo{volume}{D75}}, \bibinfo{pages}{054032}
  (\bibinfo{year}{2007}), \eprint{hep-ph/0701006}\relax
\mciteBstWouldAddEndPuncttrue
\mciteSetBstMidEndSepPunct{\mcitedefaultmidpunct}
{\mcitedefaultendpunct}{\mcitedefaultseppunct}\relax
\EndOfBibitem
\bibitem[{\citenamefont{Collins}(1993)}]{Collins:1993kk}
\bibinfo{author}{\bibfnamefont{J.~C.} \bibnamefont{Collins}},
  \bibinfo{journal}{Nucl. Phys.} \textbf{\bibinfo{volume}{B396}},
  \bibinfo{pages}{161} (\bibinfo{year}{1993}),
  \eprint[http://arXiv.org/abs]{hep-ph/9208213}\relax
\mciteBstWouldAddEndPuncttrue
\mciteSetBstMidEndSepPunct{\mcitedefaultmidpunct}
{\mcitedefaultendpunct}{\mcitedefaultseppunct}\relax
\EndOfBibitem
\bibitem[{\citenamefont{Kang et~al.}(2016)\citenamefont{Kang, Prokudin, Sun,
  and Yuan}}]{Kang:2015msa}
\bibinfo{author}{\bibfnamefont{Z.-B.} \bibnamefont{Kang}},
  \bibinfo{author}{\bibfnamefont{A.}~\bibnamefont{Prokudin}},
  \bibinfo{author}{\bibfnamefont{P.}~\bibnamefont{Sun}}, \bibnamefont{and}
  \bibinfo{author}{\bibfnamefont{F.}~\bibnamefont{Yuan}},
  \bibinfo{journal}{Phys. Rev.} \textbf{\bibinfo{volume}{D93}},
  \bibinfo{pages}{014009} (\bibinfo{year}{2016}), \eprint{1505.05589}\relax
\mciteBstWouldAddEndPuncttrue
\mciteSetBstMidEndSepPunct{\mcitedefaultmidpunct}
{\mcitedefaultendpunct}{\mcitedefaultseppunct}\relax
\EndOfBibitem
\bibitem[{\citenamefont{Rogers and Mulders}(2010)}]{Rogers:2010dm}
\bibinfo{author}{\bibfnamefont{T.~C.} \bibnamefont{Rogers}} \bibnamefont{and}
  \bibinfo{author}{\bibfnamefont{P.~J.} \bibnamefont{Mulders}},
  \bibinfo{journal}{Phys. Rev.} \textbf{\bibinfo{volume}{D81}},
  \bibinfo{pages}{094006} (\bibinfo{year}{2010}), \eprint{1001.2977}\relax
\mciteBstWouldAddEndPuncttrue
\mciteSetBstMidEndSepPunct{\mcitedefaultmidpunct}
{\mcitedefaultendpunct}{\mcitedefaultseppunct}\relax
\EndOfBibitem
\bibitem[{\citenamefont{Jaffe et~al.}(1998)\citenamefont{Jaffe, Jin, and
  Tang}}]{Jaffe:1998hf}
\bibinfo{author}{\bibfnamefont{R.~L.} \bibnamefont{Jaffe}},
  \bibinfo{author}{\bibfnamefont{X.}~\bibnamefont{Jin}}, \bibnamefont{and}
  \bibinfo{author}{\bibfnamefont{J.}~\bibnamefont{Tang}},
  \bibinfo{journal}{Phys. Rev. Lett.} \textbf{\bibinfo{volume}{80}},
  \bibinfo{pages}{1166} (\bibinfo{year}{1998}),
  \eprint[http://arXiv.org/abs]{hep-ph/9709322}\relax
\mciteBstWouldAddEndPuncttrue
\mciteSetBstMidEndSepPunct{\mcitedefaultmidpunct}
{\mcitedefaultendpunct}{\mcitedefaultseppunct}\relax
\EndOfBibitem
\bibitem[{\citenamefont{Collins and Ladinsky}(1994)}]{Collins:1994ax}
\bibinfo{author}{\bibfnamefont{J.~C.} \bibnamefont{Collins}} \bibnamefont{and}
  \bibinfo{author}{\bibfnamefont{G.~A.} \bibnamefont{Ladinsky}}
  (\bibinfo{year}{1994}), \eprint[http://arXiv.org/abs]{hep-ph/9411444}\relax
\mciteBstWouldAddEndPuncttrue
\mciteSetBstMidEndSepPunct{\mcitedefaultmidpunct}
{\mcitedefaultendpunct}{\mcitedefaultseppunct}\relax
\EndOfBibitem
\bibitem[{\citenamefont{Bianconi et~al.}(2000)\citenamefont{Bianconi, Boffi,
  Jakob, and Radici}}]{Bianconi:1999cd}
\bibinfo{author}{\bibfnamefont{A.}~\bibnamefont{Bianconi}},
  \bibinfo{author}{\bibfnamefont{S.}~\bibnamefont{Boffi}},
  \bibinfo{author}{\bibfnamefont{R.}~\bibnamefont{Jakob}}, \bibnamefont{and}
  \bibinfo{author}{\bibfnamefont{M.}~\bibnamefont{Radici}},
  \bibinfo{journal}{Phys. Rev.} \textbf{\bibinfo{volume}{D62}},
  \bibinfo{pages}{034008} (\bibinfo{year}{2000}),
  \eprint[http://arXiv.org/abs]{hep-ph/9907475}\relax
\mciteBstWouldAddEndPuncttrue
\mciteSetBstMidEndSepPunct{\mcitedefaultmidpunct}
{\mcitedefaultendpunct}{\mcitedefaultseppunct}\relax
\EndOfBibitem
\bibitem[{\citenamefont{Radici et~al.}(2002)\citenamefont{Radici, Jakob, and
  Bianconi}}]{Radici:2001na}
\bibinfo{author}{\bibfnamefont{M.}~\bibnamefont{Radici}},
  \bibinfo{author}{\bibfnamefont{R.}~\bibnamefont{Jakob}}, \bibnamefont{and}
  \bibinfo{author}{\bibfnamefont{A.}~\bibnamefont{Bianconi}},
  \bibinfo{journal}{Phys. Rev.} \textbf{\bibinfo{volume}{D65}},
  \bibinfo{pages}{074031} (\bibinfo{year}{2002}),
  \eprint[http://arXiv.org/abs]{hep-ph/0110252}\relax
\mciteBstWouldAddEndPuncttrue
\mciteSetBstMidEndSepPunct{\mcitedefaultmidpunct}
{\mcitedefaultendpunct}{\mcitedefaultseppunct}\relax
\EndOfBibitem
\bibitem[{\citenamefont{Bacchetta and Radici}(2003)}]{Bacchetta:2002ux}
\bibinfo{author}{\bibfnamefont{A.}~\bibnamefont{Bacchetta}} \bibnamefont{and}
  \bibinfo{author}{\bibfnamefont{M.}~\bibnamefont{Radici}},
  \bibinfo{journal}{Phys. Rev.} \textbf{\bibinfo{volume}{D67}},
  \bibinfo{pages}{094002} (\bibinfo{year}{2003}), \eprint{hep-ph/0212300}\relax
\mciteBstWouldAddEndPuncttrue
\mciteSetBstMidEndSepPunct{\mcitedefaultmidpunct}
{\mcitedefaultendpunct}{\mcitedefaultseppunct}\relax
\EndOfBibitem
\bibitem[{\citenamefont{Boer et~al.}(2003)\citenamefont{Boer, Jakob, and
  Radici}}]{Boer:2003ya}
\bibinfo{author}{\bibfnamefont{D.}~\bibnamefont{Boer}},
  \bibinfo{author}{\bibfnamefont{R.}~\bibnamefont{Jakob}}, \bibnamefont{and}
  \bibinfo{author}{\bibfnamefont{M.}~\bibnamefont{Radici}},
  \bibinfo{journal}{Phys. Rev.} \textbf{\bibinfo{volume}{D67}},
  \bibinfo{pages}{094003} (\bibinfo{year}{2003}), \eprint{hep-ph/0302232}\relax
\mciteBstWouldAddEndPuncttrue
\mciteSetBstMidEndSepPunct{\mcitedefaultmidpunct}
{\mcitedefaultendpunct}{\mcitedefaultseppunct}\relax
\EndOfBibitem
\bibitem[{\citenamefont{Bacchetta et~al.}(2009)\citenamefont{Bacchetta,
  Ceccopieri, Mukherjee, and Radici}}]{Bacchetta:2008wb}
\bibinfo{author}{\bibfnamefont{A.}~\bibnamefont{Bacchetta}},
  \bibinfo{author}{\bibfnamefont{F.~A.} \bibnamefont{Ceccopieri}},
  \bibinfo{author}{\bibfnamefont{A.}~\bibnamefont{Mukherjee}},
  \bibnamefont{and} \bibinfo{author}{\bibfnamefont{M.}~\bibnamefont{Radici}},
  \bibinfo{journal}{Phys. Rev.} \textbf{\bibinfo{volume}{D79}},
  \bibinfo{pages}{034029} (\bibinfo{year}{2009}), \eprint{0812.0611}\relax
\mciteBstWouldAddEndPuncttrue
\mciteSetBstMidEndSepPunct{\mcitedefaultmidpunct}
{\mcitedefaultendpunct}{\mcitedefaultseppunct}\relax
\EndOfBibitem
\bibitem[{\citenamefont{Courtoy et~al.}(2012)\citenamefont{Courtoy, Bacchetta,
  Radici, and Bianconi}}]{Courtoy:2012ry}
\bibinfo{author}{\bibfnamefont{A.}~\bibnamefont{Courtoy}},
  \bibinfo{author}{\bibfnamefont{A.}~\bibnamefont{Bacchetta}},
  \bibinfo{author}{\bibfnamefont{M.}~\bibnamefont{Radici}}, \bibnamefont{and}
  \bibinfo{author}{\bibfnamefont{A.}~\bibnamefont{Bianconi}},
  \bibinfo{journal}{Phys.Rev.} \textbf{\bibinfo{volume}{D85}},
  \bibinfo{pages}{114023} (\bibinfo{year}{2012}), \eprint{1202.0323}\relax
\mciteBstWouldAddEndPuncttrue
\mciteSetBstMidEndSepPunct{\mcitedefaultmidpunct}
{\mcitedefaultendpunct}{\mcitedefaultseppunct}\relax
\EndOfBibitem
\bibitem[{\citenamefont{Bacchetta and Radici}(2004)}]{Bacchetta:2004it}
\bibinfo{author}{\bibfnamefont{A.}~\bibnamefont{Bacchetta}} \bibnamefont{and}
  \bibinfo{author}{\bibfnamefont{M.}~\bibnamefont{Radici}},
  \bibinfo{journal}{Phys. Rev.} \textbf{\bibinfo{volume}{D70}},
  \bibinfo{pages}{094032} (\bibinfo{year}{2004}), \eprint{hep-ph/0409174}\relax
\mciteBstWouldAddEndPuncttrue
\mciteSetBstMidEndSepPunct{\mcitedefaultmidpunct}
{\mcitedefaultendpunct}{\mcitedefaultseppunct}\relax
\EndOfBibitem
\bibitem[{\citenamefont{Radici et~al.}(2016)\citenamefont{Radici, Ricci,
  Bacchetta, and Mukherjee}}]{Radici:2016lam}
\bibinfo{author}{\bibfnamefont{M.}~\bibnamefont{Radici}},
  \bibinfo{author}{\bibfnamefont{A.~M.} \bibnamefont{Ricci}},
  \bibinfo{author}{\bibfnamefont{A.}~\bibnamefont{Bacchetta}},
  \bibnamefont{and}
  \bibinfo{author}{\bibfnamefont{A.}~\bibnamefont{Mukherjee}},
  \bibinfo{journal}{Phys. Rev.} \textbf{\bibinfo{volume}{D94}},
  \bibinfo{pages}{034012} (\bibinfo{year}{2016}), \eprint{1604.06585}\relax
\mciteBstWouldAddEndPuncttrue
\mciteSetBstMidEndSepPunct{\mcitedefaultmidpunct}
{\mcitedefaultendpunct}{\mcitedefaultseppunct}\relax
\EndOfBibitem
\bibitem[{\citenamefont{Airapetian et~al.}(2008)}]{Airapetian:2008sk}
\bibinfo{author}{\bibfnamefont{A.}~\bibnamefont{Airapetian}}
  \bibnamefont{et~al.} (\bibinfo{collaboration}{HERMES}),
  \bibinfo{journal}{JHEP} \textbf{\bibinfo{volume}{06}}, \bibinfo{pages}{017}
  (\bibinfo{year}{2008}), \eprint{0803.2367}\relax
\mciteBstWouldAddEndPuncttrue
\mciteSetBstMidEndSepPunct{\mcitedefaultmidpunct}
{\mcitedefaultendpunct}{\mcitedefaultseppunct}\relax
\EndOfBibitem
\bibitem[{\citenamefont{Adolph et~al.}(2012)}]{Adolph:2012nw}
\bibinfo{author}{\bibfnamefont{C.}~\bibnamefont{Adolph}} \bibnamefont{et~al.}
  (\bibinfo{collaboration}{COMPASS}), \bibinfo{journal}{Phys.Lett.}
  \textbf{\bibinfo{volume}{B713}}, \bibinfo{pages}{10} (\bibinfo{year}{2012}),
  \eprint{1202.6150}\relax
\mciteBstWouldAddEndPuncttrue
\mciteSetBstMidEndSepPunct{\mcitedefaultmidpunct}
{\mcitedefaultendpunct}{\mcitedefaultseppunct}\relax
\EndOfBibitem
\bibitem[{\citenamefont{Adolph et~al.}(2014)}]{Adolph:2014fjw}
\bibinfo{author}{\bibfnamefont{C.}~\bibnamefont{Adolph}} \bibnamefont{et~al.}
  (\bibinfo{collaboration}{COMPASS}), \bibinfo{journal}{Phys.Lett.}
  \textbf{\bibinfo{volume}{B736}}, \bibinfo{pages}{124} (\bibinfo{year}{2014}),
  \eprint{1401.7873}\relax
\mciteBstWouldAddEndPuncttrue
\mciteSetBstMidEndSepPunct{\mcitedefaultmidpunct}
{\mcitedefaultendpunct}{\mcitedefaultseppunct}\relax
\EndOfBibitem
\bibitem[{\citenamefont{Braun}(2015)}]{Braun:2015baa}
\bibinfo{author}{\bibfnamefont{C.}~\bibnamefont{Braun}}
  (\bibinfo{collaboration}{COMPASS}), \bibinfo{journal}{EPJ Web Conf.}
  \textbf{\bibinfo{volume}{85}}, \bibinfo{pages}{02018}
  (\bibinfo{year}{2015})\relax
\mciteBstWouldAddEndPuncttrue
\mciteSetBstMidEndSepPunct{\mcitedefaultmidpunct}
{\mcitedefaultendpunct}{\mcitedefaultseppunct}\relax
\EndOfBibitem
\bibitem[{\citenamefont{Vossen et~al.}(2011)}]{Vossen:2011fk}
\bibinfo{author}{\bibfnamefont{A.}~\bibnamefont{Vossen}} \bibnamefont{et~al.}
  (\bibinfo{collaboration}{Belle Collaboration}),
  \bibinfo{journal}{Phys.Rev.Lett.} \textbf{\bibinfo{volume}{107}},
  \bibinfo{pages}{072004} (\bibinfo{year}{2011}), \eprint{1104.2425}\relax
\mciteBstWouldAddEndPuncttrue
\mciteSetBstMidEndSepPunct{\mcitedefaultmidpunct}
{\mcitedefaultendpunct}{\mcitedefaultseppunct}\relax
\EndOfBibitem
\bibitem[{\citenamefont{Bacchetta et~al.}(2011)\citenamefont{Bacchetta,
  Courtoy, and Radici}}]{Bacchetta:2011ip}
\bibinfo{author}{\bibfnamefont{A.}~\bibnamefont{Bacchetta}},
  \bibinfo{author}{\bibfnamefont{A.}~\bibnamefont{Courtoy}}, \bibnamefont{and}
  \bibinfo{author}{\bibfnamefont{M.}~\bibnamefont{Radici}},
  \bibinfo{journal}{Phys.Rev.Lett.} \textbf{\bibinfo{volume}{107}},
  \bibinfo{pages}{012001} (\bibinfo{year}{2011}), \eprint{1104.3855}\relax
\mciteBstWouldAddEndPuncttrue
\mciteSetBstMidEndSepPunct{\mcitedefaultmidpunct}
{\mcitedefaultendpunct}{\mcitedefaultseppunct}\relax
\EndOfBibitem
\bibitem[{\citenamefont{Bacchetta et~al.}(2013)\citenamefont{Bacchetta,
  Courtoy, and Radici}}]{Bacchetta:2012ty}
\bibinfo{author}{\bibfnamefont{A.}~\bibnamefont{Bacchetta}},
  \bibinfo{author}{\bibfnamefont{A.}~\bibnamefont{Courtoy}}, \bibnamefont{and}
  \bibinfo{author}{\bibfnamefont{M.}~\bibnamefont{Radici}},
  \bibinfo{journal}{JHEP} \textbf{\bibinfo{volume}{1303}}, \bibinfo{pages}{119}
  (\bibinfo{year}{2013}), \eprint{1212.3568}\relax
\mciteBstWouldAddEndPuncttrue
\mciteSetBstMidEndSepPunct{\mcitedefaultmidpunct}
{\mcitedefaultendpunct}{\mcitedefaultseppunct}\relax
\EndOfBibitem
\bibitem[{\citenamefont{Radici et~al.}(2015)\citenamefont{Radici, Courtoy,
  Bacchetta, and Guagnelli}}]{Radici:2015mwa}
\bibinfo{author}{\bibfnamefont{M.}~\bibnamefont{Radici}},
  \bibinfo{author}{\bibfnamefont{A.}~\bibnamefont{Courtoy}},
  \bibinfo{author}{\bibfnamefont{A.}~\bibnamefont{Bacchetta}},
  \bibnamefont{and}
  \bibinfo{author}{\bibfnamefont{M.}~\bibnamefont{Guagnelli}},
  \bibinfo{journal}{JHEP} \textbf{\bibinfo{volume}{05}}, \bibinfo{pages}{123}
  (\bibinfo{year}{2015}), \eprint{1503.03495}\relax
\mciteBstWouldAddEndPuncttrue
\mciteSetBstMidEndSepPunct{\mcitedefaultmidpunct}
{\mcitedefaultendpunct}{\mcitedefaultseppunct}\relax
\EndOfBibitem
\bibitem[{\citenamefont{Adamczyk et~al.}(2015)}]{Adamczyk:2015hri}
\bibinfo{author}{\bibfnamefont{L.}~\bibnamefont{Adamczyk}} \bibnamefont{et~al.}
  (\bibinfo{collaboration}{STAR}), \bibinfo{journal}{Phys. Rev. Lett.}
  \textbf{\bibinfo{volume}{115}}, \bibinfo{pages}{242501}
  (\bibinfo{year}{2015}), \eprint{1504.00415}\relax
\mciteBstWouldAddEndPuncttrue
\mciteSetBstMidEndSepPunct{\mcitedefaultmidpunct}
{\mcitedefaultendpunct}{\mcitedefaultseppunct}\relax
\EndOfBibitem
\bibitem[{\citenamefont{Matevosyan et~al.}(2018)\citenamefont{Matevosyan,
  Bacchetta, Boer, Courtoy, Kotzinian, Radici, and
  Thomas}}]{Matevosyan:2018icf}
\bibinfo{author}{\bibfnamefont{H.~H.} \bibnamefont{Matevosyan}},
  \bibinfo{author}{\bibfnamefont{A.}~\bibnamefont{Bacchetta}},
  \bibinfo{author}{\bibfnamefont{D.}~\bibnamefont{Boer}},
  \bibinfo{author}{\bibfnamefont{A.}~\bibnamefont{Courtoy}},
  \bibinfo{author}{\bibfnamefont{A.}~\bibnamefont{Kotzinian}},
  \bibinfo{author}{\bibfnamefont{M.}~\bibnamefont{Radici}}, \bibnamefont{and}
  \bibinfo{author}{\bibfnamefont{A.~W.} \bibnamefont{Thomas}},
  \bibinfo{journal}{Phys. Rev.} \textbf{\bibinfo{volume}{D97}},
  \bibinfo{pages}{074019} (\bibinfo{year}{2018}), \eprint{1802.01578}\relax
\mciteBstWouldAddEndPuncttrue
\mciteSetBstMidEndSepPunct{\mcitedefaultmidpunct}
{\mcitedefaultendpunct}{\mcitedefaultseppunct}\relax
\EndOfBibitem
\bibitem[{\citenamefont{Artru and Collins}(1996)}]{Artru:1995zu}
\bibinfo{author}{\bibfnamefont{X.}~\bibnamefont{Artru}} \bibnamefont{and}
  \bibinfo{author}{\bibfnamefont{J.~C.} \bibnamefont{Collins}},
  \bibinfo{journal}{Z.Phys.} \textbf{\bibinfo{volume}{C69}},
  \bibinfo{pages}{277} (\bibinfo{year}{1996}), \eprint{hep-ph/9504220}\relax
\mciteBstWouldAddEndPuncttrue
\mciteSetBstMidEndSepPunct{\mcitedefaultmidpunct}
{\mcitedefaultendpunct}{\mcitedefaultseppunct}\relax
\EndOfBibitem
\bibitem[{\citenamefont{Seidl et~al.}(2017)}]{Seidl:2017qhp}
\bibinfo{author}{\bibfnamefont{R.}~\bibnamefont{Seidl}} \bibnamefont{et~al.}
  (\bibinfo{collaboration}{Belle}), \bibinfo{journal}{Phys. Rev.}
  \textbf{\bibinfo{volume}{D96}}, \bibinfo{pages}{032005}
  (\bibinfo{year}{2017}), \eprint{1706.08348}\relax
\mciteBstWouldAddEndPuncttrue
\mciteSetBstMidEndSepPunct{\mcitedefaultmidpunct}
{\mcitedefaultendpunct}{\mcitedefaultseppunct}\relax
\EndOfBibitem
\bibitem[{\citenamefont{Stratmann and Vogelsang}(2001)}]{Stratmann:2001pb}
\bibinfo{author}{\bibfnamefont{M.}~\bibnamefont{Stratmann}} \bibnamefont{and}
  \bibinfo{author}{\bibfnamefont{W.}~\bibnamefont{Vogelsang}},
  \bibinfo{journal}{Phys. Rev.} \textbf{\bibinfo{volume}{D64}},
  \bibinfo{pages}{114007} (\bibinfo{year}{2001}), \eprint{hep-ph/0107064}\relax
\mciteBstWouldAddEndPuncttrue
\mciteSetBstMidEndSepPunct{\mcitedefaultmidpunct}
{\mcitedefaultendpunct}{\mcitedefaultseppunct}\relax
\EndOfBibitem
\bibitem[{sup()}]{suppl}
\bibinfo{note}{{\rm See the Supplemental Material at
  http://link.aps.org/supplemental/10.1103/PhysRevLett.120.192001 for details
  of the calculation of the matrix element for proton-proton collisions in
  Mellin space, for the fit to the Soffer bound at the starting scale, and for
  the formula of transversity evolved at LO in Mellin space.}}\relax
\mciteBstWouldAddEndPunctfalse
\mciteSetBstMidEndSepPunct{\mcitedefaultmidpunct}
{}{\mcitedefaultseppunct}\relax
\EndOfBibitem
\bibitem[{\citenamefont{Bacchetta and Radici}(2006)}]{Bacchetta:2006un}
\bibinfo{author}{\bibfnamefont{A.}~\bibnamefont{Bacchetta}} \bibnamefont{and}
  \bibinfo{author}{\bibfnamefont{M.}~\bibnamefont{Radici}},
  \bibinfo{journal}{Phys. Rev.} \textbf{\bibinfo{volume}{D74}},
  \bibinfo{pages}{114007} (\bibinfo{year}{2006}), \eprint{hep-ph/0608037}\relax
\mciteBstWouldAddEndPuncttrue
\mciteSetBstMidEndSepPunct{\mcitedefaultmidpunct}
{\mcitedefaultendpunct}{\mcitedefaultseppunct}\relax
\EndOfBibitem
\bibitem[{\citenamefont{Vogelsang}(1998)}]{Vogelsang:1997ak}
\bibinfo{author}{\bibfnamefont{W.}~\bibnamefont{Vogelsang}},
  \bibinfo{journal}{Phys. Rev.} \textbf{\bibinfo{volume}{D57}},
  \bibinfo{pages}{1886} (\bibinfo{year}{1998}), \eprint{hep-ph/9706511}\relax
\mciteBstWouldAddEndPuncttrue
\mciteSetBstMidEndSepPunct{\mcitedefaultmidpunct}
{\mcitedefaultendpunct}{\mcitedefaultseppunct}\relax
\EndOfBibitem
\bibitem[{\citenamefont{Martin et~al.}(2009)\citenamefont{Martin, Stirling,
  Thorne, and Watt}}]{Martin:2009iq}
\bibinfo{author}{\bibfnamefont{A.~D.} \bibnamefont{Martin}},
  \bibinfo{author}{\bibfnamefont{W.~J.} \bibnamefont{Stirling}},
  \bibinfo{author}{\bibfnamefont{R.~S.} \bibnamefont{Thorne}},
  \bibnamefont{and} \bibinfo{author}{\bibfnamefont{G.}~\bibnamefont{Watt}},
  \bibinfo{journal}{Eur. Phys. J.} \textbf{\bibinfo{volume}{C63}},
  \bibinfo{pages}{189} (\bibinfo{year}{2009}), \eprint{0901.0002}\relax
\mciteBstWouldAddEndPuncttrue
\mciteSetBstMidEndSepPunct{\mcitedefaultmidpunct}
{\mcitedefaultendpunct}{\mcitedefaultseppunct}\relax
\EndOfBibitem
\bibitem[{\citenamefont{de~Florian et~al.}(2009)\citenamefont{de~Florian,
  Sassot, Stratmann, and Vogelsang}}]{deFlorian:2009vb}
\bibinfo{author}{\bibfnamefont{D.}~\bibnamefont{de~Florian}},
  \bibinfo{author}{\bibfnamefont{R.}~\bibnamefont{Sassot}},
  \bibinfo{author}{\bibfnamefont{M.}~\bibnamefont{Stratmann}},
  \bibnamefont{and}
  \bibinfo{author}{\bibfnamefont{W.}~\bibnamefont{Vogelsang}},
  \bibinfo{journal}{Phys.Rev.} \textbf{\bibinfo{volume}{D80}},
  \bibinfo{pages}{034030} (\bibinfo{year}{2009}), \eprint{0904.3821}\relax
\mciteBstWouldAddEndPuncttrue
\mciteSetBstMidEndSepPunct{\mcitedefaultmidpunct}
{\mcitedefaultendpunct}{\mcitedefaultseppunct}\relax
\EndOfBibitem
\bibitem[{\citenamefont{Sato et~al.}(2016)\citenamefont{Sato, Melnitchouk,
  Kuhn, Ethier, and Accardi}}]{Sato:2016tuz}
\bibinfo{author}{\bibfnamefont{N.}~\bibnamefont{Sato}},
  \bibinfo{author}{\bibfnamefont{W.}~\bibnamefont{Melnitchouk}},
  \bibinfo{author}{\bibfnamefont{S.~E.} \bibnamefont{Kuhn}},
  \bibinfo{author}{\bibfnamefont{J.~J.} \bibnamefont{Ethier}},
  \bibnamefont{and} \bibinfo{author}{\bibfnamefont{A.}~\bibnamefont{Accardi}}
  (\bibinfo{collaboration}{Jefferson Lab Angular Momentum}),
  \bibinfo{journal}{Phys. Rev.} \textbf{\bibinfo{volume}{D93}},
  \bibinfo{pages}{074005} (\bibinfo{year}{2016}), \eprint{1601.07782}\relax
\mciteBstWouldAddEndPuncttrue
\mciteSetBstMidEndSepPunct{\mcitedefaultmidpunct}
{\mcitedefaultendpunct}{\mcitedefaultseppunct}\relax
\EndOfBibitem
\bibitem[{\citenamefont{Ethier et~al.}(2017)\citenamefont{Ethier, Sato, and
  Melnitchouk}}]{Ethier:2017zbq}
\bibinfo{author}{\bibfnamefont{J.~J.} \bibnamefont{Ethier}},
  \bibinfo{author}{\bibfnamefont{N.}~\bibnamefont{Sato}}, \bibnamefont{and}
  \bibinfo{author}{\bibfnamefont{W.}~\bibnamefont{Melnitchouk}},
  \bibinfo{journal}{Phys. Rev. Lett.} \textbf{\bibinfo{volume}{119}},
  \bibinfo{pages}{132001} (\bibinfo{year}{2017}), \eprint{1705.05889}\relax
\mciteBstWouldAddEndPuncttrue
\mciteSetBstMidEndSepPunct{\mcitedefaultmidpunct}
{\mcitedefaultendpunct}{\mcitedefaultseppunct}\relax
\EndOfBibitem
\bibitem[{\citenamefont{Accardi and Bacchetta}(2017)}]{Accardi:2017pmi}
\bibinfo{author}{\bibfnamefont{A.}~\bibnamefont{Accardi}} \bibnamefont{and}
  \bibinfo{author}{\bibfnamefont{A.}~\bibnamefont{Bacchetta}},
  \bibinfo{journal}{Phys. Lett.} \textbf{\bibinfo{volume}{B773}},
  \bibinfo{pages}{632} (\bibinfo{year}{2017}), \eprint{1706.02000}\relax
\mciteBstWouldAddEndPuncttrue
\mciteSetBstMidEndSepPunct{\mcitedefaultmidpunct}
{\mcitedefaultendpunct}{\mcitedefaultseppunct}\relax
\EndOfBibitem
\bibitem[{\citenamefont{Salam and Rojo}(2009)}]{Salam:2008qg}
\bibinfo{author}{\bibfnamefont{G.~P.} \bibnamefont{Salam}} \bibnamefont{and}
  \bibinfo{author}{\bibfnamefont{J.}~\bibnamefont{Rojo}},
  \bibinfo{journal}{Comput.Phys.Commun.} \textbf{\bibinfo{volume}{180}},
  \bibinfo{pages}{120} (\bibinfo{year}{2009}), \eprint{0804.3755}\relax
\mciteBstWouldAddEndPuncttrue
\mciteSetBstMidEndSepPunct{\mcitedefaultmidpunct}
{\mcitedefaultendpunct}{\mcitedefaultseppunct}\relax
\EndOfBibitem
\bibitem[{\citenamefont{Ceccopieri et~al.}(2007)\citenamefont{Ceccopieri,
  Radici, and Bacchetta}}]{Ceccopieri:2007ip}
\bibinfo{author}{\bibfnamefont{F.~A.} \bibnamefont{Ceccopieri}},
  \bibinfo{author}{\bibfnamefont{M.}~\bibnamefont{Radici}}, \bibnamefont{and}
  \bibinfo{author}{\bibfnamefont{A.}~\bibnamefont{Bacchetta}},
  \bibinfo{journal}{Phys. Lett.} \textbf{\bibinfo{volume}{B650}},
  \bibinfo{pages}{81} (\bibinfo{year}{2007}), \eprint{hep-ph/0703265}\relax
\mciteBstWouldAddEndPuncttrue
\mciteSetBstMidEndSepPunct{\mcitedefaultmidpunct}
{\mcitedefaultendpunct}{\mcitedefaultseppunct}\relax
\EndOfBibitem
\bibitem[{\citenamefont{Anselmino et~al.}(2013)\citenamefont{Anselmino,
  Boglione, D'Alesio, Melis, Murgia et~al.}}]{Anselmino:2013vqa}
\bibinfo{author}{\bibfnamefont{M.}~\bibnamefont{Anselmino}},
  \bibinfo{author}{\bibfnamefont{M.}~\bibnamefont{Boglione}},
  \bibinfo{author}{\bibfnamefont{U.}~\bibnamefont{D'Alesio}},
  \bibinfo{author}{\bibfnamefont{S.}~\bibnamefont{Melis}},
  \bibinfo{author}{\bibfnamefont{F.}~\bibnamefont{Murgia}},
  \bibnamefont{et~al.}, \bibinfo{journal}{Phys.Rev.}
  \textbf{\bibinfo{volume}{D87}}, \bibinfo{pages}{094019}
  (\bibinfo{year}{2013}), \eprint{1303.3822}\relax
\mciteBstWouldAddEndPuncttrue
\mciteSetBstMidEndSepPunct{\mcitedefaultmidpunct}
{\mcitedefaultendpunct}{\mcitedefaultseppunct}\relax
\EndOfBibitem
\bibitem[{\citenamefont{Radici}(2017)}]{Radici:2016opu}
\bibinfo{author}{\bibfnamefont{M.}~\bibnamefont{Radici}},
  \bibinfo{journal}{PoS} \textbf{\bibinfo{volume}{QCDEV2016}},
  \bibinfo{pages}{013} (\bibinfo{year}{2017}), \eprint{1611.03351}\relax
\mciteBstWouldAddEndPuncttrue
\mciteSetBstMidEndSepPunct{\mcitedefaultmidpunct}
{\mcitedefaultendpunct}{\mcitedefaultseppunct}\relax
\EndOfBibitem
\bibitem[{\citenamefont{Bhattacharya et~al.}(2015)\citenamefont{Bhattacharya,
  Cirigliano, Cohen, Gupta, Joseph, Lin, and Yoon}}]{Bhattacharya:2015wna}
\bibinfo{author}{\bibfnamefont{T.}~\bibnamefont{Bhattacharya}},
  \bibinfo{author}{\bibfnamefont{V.}~\bibnamefont{Cirigliano}},
  \bibinfo{author}{\bibfnamefont{S.}~\bibnamefont{Cohen}},
  \bibinfo{author}{\bibfnamefont{R.}~\bibnamefont{Gupta}},
  \bibinfo{author}{\bibfnamefont{A.}~\bibnamefont{Joseph}},
  \bibinfo{author}{\bibfnamefont{H.-W.} \bibnamefont{Lin}}, \bibnamefont{and}
  \bibinfo{author}{\bibfnamefont{B.}~\bibnamefont{Yoon}}
  (\bibinfo{collaboration}{PNDME}), \bibinfo{journal}{Phys. Rev.}
  \textbf{\bibinfo{volume}{D92}}, \bibinfo{pages}{094511}
  (\bibinfo{year}{2015}), \eprint{1506.06411}\relax
\mciteBstWouldAddEndPuncttrue
\mciteSetBstMidEndSepPunct{\mcitedefaultmidpunct}
{\mcitedefaultendpunct}{\mcitedefaultseppunct}\relax
\EndOfBibitem
\bibitem[{\citenamefont{Lin et~al.}(2018)\citenamefont{Lin, Melnitchouk,
  Prokudin, Sato, and Shows}}]{Lin:2017stx}
\bibinfo{author}{\bibfnamefont{H.-W.} \bibnamefont{Lin}},
  \bibinfo{author}{\bibfnamefont{W.}~\bibnamefont{Melnitchouk}},
  \bibinfo{author}{\bibfnamefont{A.}~\bibnamefont{Prokudin}},
  \bibinfo{author}{\bibfnamefont{N.}~\bibnamefont{Sato}}, \bibnamefont{and}
  \bibinfo{author}{\bibfnamefont{H.}~\bibnamefont{Shows}},
  \bibinfo{journal}{Phys. Rev. Lett.} \textbf{\bibinfo{volume}{120}},
  \bibinfo{pages}{152502} (\bibinfo{year}{2018}), \eprint{1710.09858}\relax
\mciteBstWouldAddEndPuncttrue
\mciteSetBstMidEndSepPunct{\mcitedefaultmidpunct}
{\mcitedefaultendpunct}{\mcitedefaultseppunct}\relax
\EndOfBibitem
\bibitem[{\citenamefont{Cirigliano et~al.}(2013)\citenamefont{Cirigliano,
  Gardner, and Holstein}}]{Cirigliano:2013xha}
\bibinfo{author}{\bibfnamefont{V.}~\bibnamefont{Cirigliano}},
  \bibinfo{author}{\bibfnamefont{S.}~\bibnamefont{Gardner}}, \bibnamefont{and}
  \bibinfo{author}{\bibfnamefont{B.}~\bibnamefont{Holstein}},
  \bibinfo{journal}{Prog.Part.Nucl.Phys.} \textbf{\bibinfo{volume}{71}},
  \bibinfo{pages}{93} (\bibinfo{year}{2013}), \eprint{1303.6953}\relax
\mciteBstWouldAddEndPuncttrue
\mciteSetBstMidEndSepPunct{\mcitedefaultmidpunct}
{\mcitedefaultendpunct}{\mcitedefaultseppunct}\relax
\EndOfBibitem
\end{mcitethebibliography}


\end{document}